\documentclass[amsmath,amssymb,pra,showpacs,twocolumn]{revtex4}

\begin{document}

\title{On entanglement at tripartite states}

\author{\framebox{S. Bugajski}, J. A. Miszczak\footnote{e-mail: miszczak@iitis.gliwice.pl}}
\affiliation{Institute of Physics of the University of Silesia, Katowice}
\affiliation{Institute of Pure and Applied Informatics of the Polish Academy of Sciences, Gliwice}

\date{\today}

\begin{abstract}
Article presents general formulation of entanglement measures problem in terms of correlation function. Description of entanglement in probabilistic framework allow us to introduce new quantity which describes quantum and classical correlations. This formalism is applied to calculate bipartite and tripartite correlations in two special cases of entangled states of tripartite systems.
\end{abstract}

\maketitle
\section{Introduction}

Entanglement \cite{werner:epr} is actually one of the most intensively studied concepts of quantum mechanics. Current efforts are concentrated mainly on two closely related problems: find necessary and sufficient conditions for inseparability (criteria of entanglement), and find reliable quantification of entanglement (measures of entanglement). While both problems are considered solved for bi-partie quantum systems, they seem to be open for $n $-component systems with $n>2$ (see \cite{fundamentals:keyl}, \cite{terhal:detecting}, \cite{bruss:characterizing} for a review and references). It is clear, however, that none of the mentioned problems could be solved without answering the more fundamental question: how to separate classical and quantum correlations?

Recently a new approach to entanglement has been proposed
\cite{beltrametti:correlations,beltrametti:entanglement}. It is
based on two assumptions: \textit{entanglement is a~kind of
correlation between values of jointly measured observables}, and:
\textit{mixing does not produce entanglement}. Important
consequences of these assumptions are: a natural discrimination
between classical and quantum correlations, and a complete
description of both kinds of correlation in terms of density
functions. Although we believe that the concept of quantum
correlation captures exactly the quantum phenomenon of
entanglement, we prefer to use the term ''quantum correlation''
instead of ''entanglement'', because the latter is understood in
many different and, probably, inequivalent ways.

The first of the two mentioned problems finds in this approach a~natural solution. As the appropriate density function $\phi_{q}$ (called the \textit{quantum correlation function}) provides a~complete description of quantum correlation, the most precise criterion for the presence of quantum correlation is simply
\begin{equation}
\phi_{q}\neq const.
\end{equation}
In \cite{beltrametti:entanglement} some simple cases of bipartite quantum correlations are calculated, nevertheless the approach can be easily generalized to $n$-component systems with arbitrary~$n$. Here we calculate quantum and classical correlations of some typical quantum observables at the tripartite GHZ state and at the W state \cite{drull:three}. A~study of tripartite correlations become especially important since the experimental realization of the GHZ state \cite{nelson:experimental}.

\section{Quantum joint observables}
\subsection{Joint observables in terms of POV measures}

In modern quantum mechanics, observables (experimentally measurable properties) are represented by POV (Positive Operator Valued) measures. Here we will be concerned with discrete observables only, their formal representation is even simpler. Let $\Xi=\left\{\xi_{1},\xi_{2},....\right\}$ be a finite or countable infinite set of possible values of a~discrete observable. The quantum mechanical representation of such an observable is provided by a~\emph{POV function} $E$ which to every possible value $\xi_{i}$ associates a self-adjoint operator $E\left(\xi_{i}\right)$ on the underlying Hilbert space $\mathcal{H}$; all the operators $E\left(\xi_{i}\right)$ have to satisfy the following two conditions:
\begin{equation}
0\leqslant\left\langle \psi\left|  E\left(  \xi_{i}\right)  \right|
\psi\right\rangle \leqslant1
\quad \mathrm{and}\quad
\sum_{i}\left\langle \psi\left|  E\left(  \xi_{i}\right)  \right|
\psi\right\rangle=1
\end{equation}
for every normalized vector $\left|\psi\right\rangle \in\mathcal{H}$. It is evident that any POV function defines a~POV measure (and \textit{vice versa}).

An important special case appears when the considered observable is real-valued, $\Xi\subset\mathbb{R}$, and -- moreover -- all operators $E\left( \xi_{i}\right)$ are projections -- $E\left(\xi_{i}\right)^{2}=E\left(\xi_{i}\right)$ for every $i$. POV functions having these properties will be called \emph{PV functions}. Then the sum $\sum_{i}\xi_{i}E\left(\xi_{i}\right)$ is a self-adjoint operator on $\mathcal{H}$, and we get the textbook representation of quantum observables.

Given a~density operator $\widehat{\rho}$ representing a~quantum state (pure or mixed), the discrete observable (represented by the POV function) $E$ with values $\left\{\xi_{1},\xi_{2},....\right\}$ defines the real function
$P_{E,\widehat{\rho}}:\left\{\xi_{1},\xi_{2},\ldots\right\}\rightarrow\left[0,1\right]$
\begin{equation}
P_{E,\widehat{\rho}}\left(\xi_{i}\right)=\mathrm{Tr}\left(E\left(\xi_{i}\right)\widehat{\rho}\right). \label{PF}
\end{equation}
The \emph{probability function} $P_{E,\widehat{\rho}}$, which determines the unique probability measure on the value set $\left\{\xi_{1},\xi_{2},\ldots\right\}$, provides the connecting bridge between the mathematical formalism and the physical reality: the probability measure emerging from $P_{E,\widehat{\rho}}$ is interpreted as the \emph{outcome measure} resulting from a measurement of the observable represented by $E$ at the state (represented by) $\widehat{\rho}$.

One of advantages of representing quantum observables as POV measures is a~natural way of describing joint observables. Let a~POV function $E$ represent a~discrete quantum observable such that its value set $\Xi$ has the product structure $\Xi=\Xi^{\prime}\times\Xi^{\prime\prime}$ with $\Xi^{\prime}=\left\{\xi_{1}^{\prime},\xi_{2}^{\prime},\ldots\right\}$,
$\Xi^{\prime\prime}=\left\{\xi_{1}^{\prime\prime},\xi_{2}^{\prime\prime},\ldots\right\}$, so a typical element of $\Xi$ is a pair $\left(\xi_{i}^{\prime},\xi_{j}^{\prime\prime}\right)$. Given such a POV function, we can define the POV function $E^{\prime}$ with the value set $\Xi^{\prime}$
\begin{equation}
E^{\prime}\left(\xi_{i}^{\prime}\right)=\sum_{j}E\left(\xi_{i}^{\prime},\xi_{j}^{\prime\prime}\right),
\end{equation}
as well as the POV function $E^{\prime\prime}$ with the value set $\Xi^{\prime\prime}$
\begin{equation}
E^{\prime\prime}\left(\xi_{j}^{\prime\prime}\right)=\sum_{i}E\left(\xi_{i}^{\prime},\xi_{j}^{\prime\prime}\right).
\end{equation}
The two derived POV functions represent some quantum observables, and the observable $E$ is then the \emph{joint observable} of $E^{\prime}$ and $E^{\prime\prime}$. Clearly, the probability function $P_{E^{\prime},\widehat{\rho}}:\left\{\xi_{1}^{\prime},\xi_{2}^{\prime},\ldots\right\}
\rightarrow\left[0,1\right]$ can be calculated from the probability function $P_{E,\widehat{\rho}}$
\begin{equation}
P_{E^{\prime},\widehat{\rho}}\left(\xi_{i}^{\prime}\right)=\sum_{j}P_{E,\widehat{\rho}}\left(  \xi_{i}^{\prime},\xi_{j}^{\prime\prime}\right).
\end{equation}
Needless to say, for arbitrary two discrete quantum observables the existence of their joint observable is not guaranteed.
\section{Examples of joint observables}
The concept of joint observable defined above for two observables can be easily generalized for an arbitrary finite number of observables. We provide below two examples showing joint observables for triplets of spin observables. Examples refer to a~tripartite quantum system described in terms of the tensor product space $\mathbb{C}^{2}\otimes\mathbb{C}^{2}\otimes\mathbb{C}^{2}$ (a three-qubit system).
\subsection{Case 1}
Take the following three mutually comeasurable observables
\begin{equation}
\begin{array}[c]{c}
\widehat{S}_{z,1}:=\widehat{s}_{z}\otimes\widehat{1}\otimes\widehat{1},\\
\widehat{S}_{z,2}:=\widehat{1}\otimes\widehat{s}_{z}\otimes\widehat{1},\\
\widehat{S}_{z,3}:=\widehat{1}\otimes\widehat{1}\otimes\widehat{s}_{z},
\end{array}
\end{equation}
where
\begin{equation}
\widehat{s}_{z}
:=\frac{1}{2}\left|0\right\rangle\left\langle0\right|-\frac{1}{2}\left|1\right\rangle\left\langle1\right|
\end{equation}
is the one-particle operator representing the $z$-th component of spin $\frac{1}{2}$.
The comeasurability of the three observables $\widehat{S}_{z,1}$, $\widehat
{S}_{z,2}$, $\widehat{S}_{z,3}$ implies the existence of their joint observable represented by the following PV function:
\begin{equation}
\begin{array}[c]{c}
E^{A}\left(\frac{1}{2},\frac{1}{2},\frac{1}{2}\right)=\left|000\right\rangle\left\langle 000\right|,\\
E^{A}\left(\frac{1}{2},\frac{1}{2},-\frac{1}{2}\right)=\left|001\right\rangle\left\langle 001\right|,\\
E^{A}\left(\frac{1}{2},-\frac{1}{2},\frac{1}{2}\right)=\left|010\right\rangle\left\langle 010\right|,\\
E^{A}\left(\frac{1}{2},-\frac{1}{2},-\frac{1}{2}\right)=\left|011\right\rangle\left\langle 011\right|,\\
E^{A}\left(-\frac{1}{2},\frac{1}{2},\frac{1}{2}\right)=\left|100\right\rangle\left\langle 100\right|,\\
E^{A}\left(-\frac{1}{2},\frac{1}{2},-\frac{1}{2}\right)=\left|101\right\rangle\left\langle 101\right|,\\
E^{A}\left(-\frac{1}{2},-\frac{1}{2},\frac{1}{2}\right)=\left|110\right\rangle\left\langle 110\right|,\\
E^{A}\left(-\frac{1}{2},-\frac{1}{2},-\frac{1}{2}\right)=\left|111\right\rangle\left\langle 111\right|.
\end{array}\label{A}
\end{equation}
\subsection{Case 2}
Let us take now
\begin{equation}
\begin{array}[c]{c}
\widehat{S}_{x,1}:=\widehat{s}_{x}\otimes\widehat{1}\otimes\widehat{1},\\
\widehat{S}_{y,2}:=\widehat{1}\otimes\widehat{s}_{y}\otimes\widehat{1},\\
\widehat{S}_{y,3}:=\widehat{1}\otimes\widehat{1}\otimes\widehat{s}_{y},
\end{array}
\end{equation}
where $\widehat{s}_{x}$ and $\widehat{s}_{y}$ are the operators of components of spin $\frac{1}{2}$. Their spectral resolutions are:
\begin{eqnarray}
\widehat{s}_{x}=\frac{1}{2}\widehat{P}_{x}-\frac{1}{2}\widehat{P}_{-x},\\
\widehat{s}_{y}=\frac{1}{2}\widehat{P}_{y}-\frac{1}{2}\widehat{P}_{-y},
\end{eqnarray}
with $\widehat{P}_{-x}=\widehat{1}-\widehat{P}_{x}$,
$\widehat{P}_{-y}=\widehat{1}-\widehat{P}_{y}$ and the projection operators $\widehat{P}_{x}$ and $\widehat{P}_{y}$ are defined by:
\begin{eqnarray}
\widehat{P}_{x}\left|0\right\rangle & =\widehat{P}_{x}\left|1\right\rangle =\frac{1}{2}\left(\left|0\right\rangle +\left|1\right\rangle
\right) \\
\widehat{P}_{y}\left|0\right\rangle  & =\frac{1}{2}\left(\left|0\right\rangle +i\left|1\right\rangle \right)  ,\;\widehat{P}_{y}\left|1\right\rangle =\frac{1}{2}\left(  -i\left|0\right\rangle +\left|1\right\rangle \right)
\end{eqnarray}

The joint observable for the three spin observables $\widehat{S}_{x,1},\widehat{S}_{y,2},\widehat{S}_{y,3}$ is represented by the following PV
function:
\begin{equation}
\begin{array}[c]{c}
E^{B}\left(\frac{1}{2},\frac{1}{2},\frac{1}{2}\right)=\widehat{P}_{x}\otimes\widehat{P}_{y}\otimes\widehat{P}_{y},\\
E^{B}\left(\frac{1}{2},\frac{1}{2},-\frac{1}{2}\right)=\widehat{P}_{x}\otimes\widehat{P}_{y}\otimes\widehat{P}_{-y},\\
E^{B}\left(\frac{1}{2},-\frac{1}{2},\frac{1}{2}\right)=\widehat{P}_{x}\otimes\widehat{P}_{-y}\otimes\widehat{P}_{y},\\
E^{B}\left(\frac{1}{2},-\frac{1}{2},-\frac{1}{2}\right)=\widehat{P}_{x}\otimes\widehat{P}_{-y}\otimes\widehat{P}_{-y},\\
E^{B}\left(-\frac{1}{2},\frac{1}{2},\frac{1}{2}\right)=\widehat{P}_{-x}\otimes\widehat{P}_{y}\otimes\widehat{P}_{y},\\
E^{B}\left(-\frac{1}{2},\frac{1}{2},-\frac{1}{2}\right)=\widehat{P}_{-x}\otimes\widehat{P}_{y}\otimes\widehat{P}_{-y},\\
E^{B}\left(-\frac{1}{2},-\frac{1}{2},\frac{1}{2}\right)=\widehat{P}_{-x}\otimes\widehat{P}_{-y}\otimes\widehat{P}_{y},\\
E^{B}\left(-\frac{1}{2},-\frac{1}{2},-\frac{1}{2}\right)=\widehat{P}_{-x}\otimes\widehat{P}_{-y}\otimes\widehat{P}_{-y}.
\label{EB}
\end{array}
\end{equation}
\section{Classical and quantum correlations}
\subsection{Total correlation}

Consider a mixed state $\widehat{\rho}$ of a three-qubit system. Assume that we are interested in describing correlations of the three observables mentioned above as Case 1 at this state. The correlation we are interested in has to be encoded in the result of a measurement of the joint observable $E^{A}$ (see 2.2.1) at $\widehat{\rho}$, so in the probability function
\begin{equation}
P_{A,\widehat{\rho}}:\left\{\frac{1}{2},-\frac{1}{2}\right\}\times\left\{\frac{1}{2},-\frac{1}{2}\right\}\times\left\{\frac{1}{2},-\frac{1}{2}\right\}\rightarrow\left[0,1\right].
\end{equation}
The function $P_{A,\widehat{\rho}}$ can be calculated according to standard rules of quantum mechanics, see Eq.~(\ref{PF}).

A measurement of the joint observable $E^{A}$ is a~joint measurement of the three observables $\widehat{S}_{z,1},$ $\widehat{S}_{z,2}$, $\widehat{S}_{z,3},$ so provides us with results of separate measurements of $\widehat{S}_{z,1},$ $\widehat{S}_{z,2}$, and $\widehat{S}_{z,3}$ at $\widehat{\rho}$. In terms of probability functions, it means that we can calculate the three functions $P_{\widehat{S}_{z,i},\widehat{\rho}}:\left\{\frac{1}{2},-\frac{1}{2}\right\}\rightarrow\left[0,1\right]$, $i=1,2,3,$ given the function $P_{A,\widehat{\rho}}$. The three probability functions $P_{\widehat{S}_{z,i},\widehat{\rho}}$ can be also calculated directly according to Eq.~(\ref{PF}).

If the three observables $\widehat{S}_{z,1},$ $\widehat{S}_{z,2}$, $\widehat{S}_{z,3}$ would be independent (uncorrelated) when jointly measured at $\widehat{\rho},$ the probability function $P_{A,\widehat{\rho}}$ should
have the product form:
\begin{equation}
P_{\widehat{S}_{z,1},\widehat{\rho}}\left(\xi_{i}^{\prime}\right)\cdot
P_{\widehat{S}_{z,2},\widehat{\rho}}\left(\xi_{j}^{\prime\prime}\right)
\cdot P_{\widehat{S}_{z,3},\widehat{\rho}}\left(\xi_{k}^{\prime\prime\prime}\right). \label{PPF}
\end{equation}
Thus, the correlations of $\widehat{S}_{z,1}$, $\widehat{S}_{z,2}$, $\widehat{S}_{z,3}$ are encoded in the ''difference'' between the product probability function and the probability function generated by the quantum joint observable $E^{A}$ at $\widehat{\rho}$. This ''difference'' is completely described by the density function $\phi_{t}:\left\{\frac
12,-\frac{1}{2}\right\}^{3}\rightarrow\left[0,1\right]$
\begin{equation}
\phi_{t}\left(\xi_{i}^{\prime},\xi_{j}^{\prime\prime},\xi_{k}^{\prime\prime\prime}\right)=\frac{P_{A,\widehat{\rho}}\left(\xi_{i}^{\prime},\xi_{j}^{\prime\prime},\xi_{k}^{\prime\prime\prime}\right)}{P_{\widehat{S}_{z,1},\widehat{\rho}}\left(\xi_{i}^{\prime}\right)\cdot P_{\widehat{S}_{z,2},\widehat{\rho}}\left(\xi_{j}^{\prime\prime}\right)\cdot
P_{\widehat{S}_{z,3},\widehat{\rho}}\left(\xi_{k}^{\prime\prime\prime}\right)}\label{TCF}
\end{equation}
Clearly, the function $\phi_{t}$ refers to total correlation, which includes both classical and quantum ones; therefore we will call it the \emph{total correlation function} (of the three observables $\widehat{S}_{z,1}$, $\widehat{S}_{z,2}$, $\widehat{S}_{z,3}$ at the state $\widehat{\rho}$). Our task now is to separate correlations of both kinds.

\subsection{Separating classical and quantum correlations}

It is evident that correlations appearing in classical probability theory (classical correlations) are caused exclusively by mixed states, and essentially depend on the way the pure states are mixed together (see \cite{beltrametti:correlations}). Thus, in the considered quantum-mechanical framework we could identify classical correlation only if we fix a ''statistical content'' of the quantum state $\widehat{\rho}$, that is if we fix a~particular decomposition of $\widehat{\rho}$ into a mixture of pure states.

Thus, assume that the considered state $\widehat{\rho}$ is a mixture of pure states $\left| \psi_{m}\right\rangle\in\mathbb{C}^{2}\otimes\mathbb{C}^{2}\otimes\mathbb{C}^{2}$
\begin{equation}
\widehat{\rho}=\sum_{m}\lambda_{m}\left|\psi_{m}\right\rangle\left\langle\psi_{m}\right|, \label{SC}
\end{equation}
where $0\leqslant\lambda_{m}$ and $\sum_{m}\lambda_{m}=1$. Any of the three considered observables $\widehat{S}_{z,1},$ $\widehat{S}_{z,2}$, $\widehat{S}_{z,3}$ defines the probability function at every pure state $\left| \psi_{m}\right\rangle$
\begin{equation}
\begin{array}[c]{c}
P_{\widehat{S}_{z,1},m}\left(\frac{1}{2}\right)=\left\langle \psi_{m}\right|
\left(\left|0\right\rangle \left\langle 0\right|\otimes\widehat
{1}\otimes\widehat{1}\right)\left|\psi_{m}\right\rangle,\\
P_{\widehat{S}_{z,1},m}\left(-\frac{1}{2}\right)=\left\langle \psi_{m}\right|
\left(\left|1\right\rangle \left\langle 1\right|\otimes\widehat
{1}\otimes\widehat{1}\right)\left|\psi_{m}\right\rangle,
\end{array}
\end{equation}
\textit{etc}. The product probability function
\begin{equation}
P_{\widehat{S}_{z,1},m}\left(\xi_{i}^{\prime}\right)P_{\widehat{S}_{z,2},m}\left(  \xi_{j}^{\prime\prime}\right)  P_{\widehat{S}_{z,3},m}\left(\xi_{k}^{\prime\prime\prime}\right)
\end{equation}
would describe the (hypothetical) situation of no correlation among $\widehat{S}_{z,1}$, $\widehat{S}_{z,2}$, $\widehat{S}_{z,3}$ at the pure state $\left|\psi_{m}\right\rangle$. Consequently, the sum
\begin{equation}
\sum_{m}\lambda_{m}P_{\widehat{S}_{z,1},m}\left(  \xi_{i}^{\prime}\right)
P_{\widehat{S}_{z,2},m}\left(  \xi_{j}^{\prime\prime}\right)  P_{\widehat
{S}_{z,3},m}\left(  \xi_{k}^{\prime\prime\prime}\right) \label{SPF}
\end{equation}
would describe only classical correlation among $\widehat{S}_{z,1}$,$\widehat{S}_{z,2}$, $\widehat{S}_{z,3}$, the one implied by the statistical content of the mixed state $\widehat{\rho}$. Assuming this (see \cite{beltrametti:correlations}, \cite{beltrametti:entanglement}), and taking into account that the product probability function, Eq.~(\ref{PPF}), contains no correlations at all, we can say that a ''difference'' between the sum probability function, Eq.~(\ref{SPF}), and the product probability function describes the classical correlation of the three observables at the mixed state $\widehat{\rho}$, Eq.~(\ref{SC}). Like the case of the total correlation, Eq.~(\ref{TCF}), the classical correlation (of observables $\widehat{S}_{z,1},$ $\widehat{S}_{z,2}$, $\widehat{S}_{z,3}$ at the mixed state $\widehat{\rho}=\sum_{m}\lambda_{m}\left|  \psi_{m}\right\rangle \left\langle \psi_{m}\right|$) is exhaustively described by the \emph{classical correlation function}: $\phi_{c}:\left\{\frac{1}{2},-\frac{1}{2}\right\}^{3}\rightarrow\left[0,1\right]$,
\begin{equation}
\phi_{c}\left(  \xi_{i}^{\prime},\xi_{j}^{\prime\prime},\xi_{k}^{\prime
\prime\prime}\right)=\frac{\sum_{m}\lambda_{m}P_{\widehat{S}_{z,1},m}\left(
\xi_{i}^{\prime}\right)  P_{\widehat{S}_{z,2},m}\left(  \xi_{j}^{\prime\prime
}\right)  P_{\widehat{S}_{z,3},m}\left(  \xi_{k}^{\prime\prime\prime}\right)
}{P_{\widehat{S}_{z,1},\widehat{\rho}}\left(  \xi_{i}^{\prime}\right)  \cdot
P_{\widehat{S}_{z,2},\widehat{\rho}}\left(  \xi_{j}^{\prime\prime}\right)
\cdot P_{\widehat{S}_{z,3},\widehat{\rho}}\left(  \xi_{k}^{\prime\prime\prime
}\right)  }\label{CCF}
\end{equation}
If the sum probability function, Eq.~(\ref{SPF}), contains exclusively classical correlations, the ''difference'' between it and the probability function $P_{A,\widehat{\rho}}$ should be caused by quantum correlations. Thus, the density function, which will be called the \emph{quantum correlation function}, $\phi_{q}:\left\{\frac{1}{2},-\frac{1}{2}\right\}^{3}\rightarrow\left[0,1\right]$,

\begin{equation}
\phi_{q}\left(\xi_{i}^{\prime},\xi_{j}^{\prime\prime},\xi_{k}^{\prime\prime\prime}\right)=\frac{P_{A,\widehat{\rho}}\left(\xi_{i}^{\prime},\xi_{j}^{\prime\prime},\xi_{k}^{\prime\prime\prime}\right)}{\sum_{m}
\lambda_{m}P_{\widehat{S}_{z,1},m}\left(\xi_{i}^{\prime}\right)P_{\widehat{S}_{z,2},m}\left(\xi_{j}^{\prime\prime}\right)P_{\widehat{S}_{z,3},m}\left(\xi_{k}^{\prime\prime\prime}\right)}
\label{QCF}
\end{equation}
should provide the full description of quantum correlations of $\widehat{S}_{z,1},$ $\widehat{S}_{z,2}$, $\widehat{S}_{z,3}$ at the mixed state $\widehat{\rho}=\sum_{m}\lambda_{m}\left|\psi_{m}\right\rangle \left\langle
\psi_{m}\right|$.

In a special case of $\widehat{\rho}=\left|\psi\right\rangle \left\langle\psi\right|  $ (a pure state), the classical correlation function becomes constant what indicates no classical correlations. In this case $\phi_{t}=\phi_{q}$, so all correlations are quantal.

Notice that the three correlation functions are connected together by the simple product rule:
\begin{equation}
\phi_{t}=\phi_{c}\cdot\phi_{q}\;.\label{PR}%
\end{equation}

\section{Quantum correlations at a GHZ state}

We are going to calculate the quantum correlation function for the two mentioned triplets of local spin observables at the GHZ state (for Greenberger, Horne, and Zeilinger \cite{ghz:goingbeyond}) of the form:
\begin{equation}
\left|  GHZ\right\rangle =\frac1{\sqrt{2}}\left(  \left|000\right\rangle
-\left|111\right\rangle \right)
\end{equation}
where $\left|0\right\rangle ,\left|1\right\rangle \in\mathbb{C}^{2}$ are eigenstates of $\widehat{s}_{z}$ corresponding to $\frac{1}{2}$ and $-\frac{1}{2}$ resp. The classical correlation function has to be constant because of purity
of the state in question.

\subsection{Case 1}

The outcome measure for the joint observable $E^{A}$ at the state $\left|GHZ\right\rangle $ is determined by the probability function $P_{A,\left|GHZ\right\rangle }$ . Standard calculations leads to:
\begin{equation}
\begin{array}{cc}
P_{A,\left|GHZ\right\rangle}\left(\frac{1}{2},\frac{1}{2},\frac{1}{2}\right)&=\frac{1}{2}\\
P_{A,\left|GHZ\right\rangle}\left(\frac{1}{2},\frac{1}{2},-\frac{1}{2}\right)&=0\\
P_{A,\left|GHZ\right\rangle}\left(\frac{1}{2},-\frac{1}{2},\frac{1}{2}\right)&=0\\
P_{A,\left|GHZ\right\rangle}\left(\frac{1}{2},-\frac{1}{2},-\frac{1}{2}\right)&=0\\
P_{A,\left|GHZ\right\rangle}\left(-\frac{1}{2},\frac{1}{2},\frac{1}{2}\right)&=0\\
P_{A,\left|GHZ\right\rangle}\left(-\frac{1}{2},\frac{1}{2},-\frac{1}{2}\right)&=0\\
P_{A,\left|GHZ\right\rangle}\left(-\frac{1}{2},-\frac{1}{2},\frac{1}{2}\right)&=0\\
P_{A,\left|GHZ\right\rangle}\left(-\frac{1}{2},-\frac{1}{2},-\frac{1}{2}\right)&=\frac{1}{2}
\end{array}
\end{equation}

The outcome measures for the three observables $\widehat{S}_{z,1}$, $\widehat{S}_{z,2}$,$\widehat{S}_{z,3}$ at
$\left|GHZ\right\rangle$ are determined by the probability functions:
\begin{equation}
\begin{array}{c}
P_{\widehat{S}_{z,i},\left|GHZ\right\rangle }\left(  \frac{1}{2}\right)=\frac{1}{2}\\
P_{\widehat{S}_{z,i},\left|GHZ\right\rangle }\left(-\frac{1}{2}\right)=\frac{1}{2}
\end{array}
\end{equation}
for $i=1,2,3,$ so the product probability function is constant:
\begin{equation}
P_{\widehat{S}_{z,1},\left|GHZ\right\rangle }\left(\xi_{i}^{\prime}\right)\cdot P_{\widehat{S}_{z,2},\left|GHZ\right\rangle }\left(\xi_{j}^{\prime\prime}\right)  \cdot P_{\widehat{S}_{z,3},\left|GHZ\right\rangle }\left(  \xi_{k}^{\prime\prime\prime}\right)=\frac18.
\end{equation}
Finally, the quantum correlation function calculated according to Eq.~(\ref{QCF}) is:
\begin{equation}
\begin{array}{c}
\phi_{q}\left(\frac{1}{2},\frac{1}{2},\frac{1}{2}\right) =4\\
\phi_{q}\left(\frac{1}{2},\frac{1}{2},-\frac{1}{2}\right) =0\\
\phi_{q}\left(\frac{1}{2},-\frac{1}{2},\frac{1}{2}\right) =0\\
\phi_{q}\left(\frac{1}{2},-\frac{1}{2},-\frac{1}{2}\right) =0\\
\phi_{q}\left(-\frac{1}{2},\frac{1}{2},\frac{1}{2}\right) =0\\
\phi_{q}\left(-\frac{1}{2},\frac{1}{2},-\frac{1}{2}\right) =0\\
\phi_{q}\left(-\frac{1}{2},-\frac{1}{2},\frac{1}{2}\right) =0\\
\phi_{q}\left(-\frac{1}{2},-\frac{1}{2},-\frac{1}{2}\right) =4
\end{array}
\end{equation}
The concentration of the obtained quantum correlation function at two points $\left(\frac{1}{2},\frac{1}{2},\frac{1}{2}\right)$ and $\left(-\frac{1}{2},-\frac{1}{2},-\frac{1}{2}\right)$ indicates that the three observables $\widehat{S}_{z,1}$, $\widehat{S}_{z,2}$, $\widehat{S}_{z,3}$ are strongly quantum correlated at $\left|GHZ\right\rangle$.

\subsection{Case 2}
The joint observable $E^{B}$, Eq.~(\ref{EB}), generates at $\left|GHZ\right\rangle$ the following outcome probability function
\begin{equation}
\begin{array}[c]{c}
P_{B,\left|GHZ\right\rangle }\left(\frac{1}{2},\frac{1}{2},\frac{1}{2}\right)=\frac{1}{4},\\
P_{B,\left|GHZ\right\rangle }\left(\frac{1}{2},\frac{1}{2},-\frac{1}{2}\right)=0,\\
P_{B,\left|GHZ\right\rangle }\left(\frac{1}{2},-\frac{1}{2},\frac{1}{2}\right)=0,\\
P_{B,\left|GHZ\right\rangle }\left(\frac{1}{2},-\frac{1}{2},-\frac{1}{2}\right)=\frac{1}{4},\\
P_{B,\left|GHZ\right\rangle }\left(-\frac{1}{2},\frac{1}{2},\frac{1}{2}\right)=0,\\
P_{B,\left|GHZ\right\rangle }\left(-\frac{1}{2},\frac{1}{2},-\frac{1}{2}\right)=\frac{1}{4},\\
P_{B,\left|GHZ\right\rangle }\left(-\frac{1}{2},-\frac{1}{2},\frac{1}{2}\right)=\frac{1}{4},\\
P_{B,\left|GHZ\right\rangle }\left(-\frac{1}{2},-\frac{1}{2},-\frac{1}{2}\right)=0.
\end{array}
\end{equation}
The outcome measures for the three observables $\widehat{S}_{x,1},\widehat {S}_{y,2},\widehat{S}_{y,3}$ at $\left|GHZ\right\rangle $ are uniformly distributed
\begin{equation}
\begin{array}[c]{c}
P_{\widehat{S}_{x,1},\left|GHZ\right\rangle }\left(\frac{1}{2}\right)=\frac{1}{2}\\
P_{\widehat{S}_{x,1},\left|GHZ\right\rangle }\left(-\frac{1}{2}\right)=\frac{1}{2}
\end{array}
\end{equation}
Hence, the product probability function is constant:
\begin{equation}
P_{\widehat{S}_{z,1},\left|GHZ\right\rangle }\left(\xi_{i}^{\prime}\right)  \cdot P_{\widehat{S}_{z,2},\left|GHZ\right\rangle }\left(\xi_{j}^{\prime\prime}\right)  \cdot P_{\widehat{S}_{z,3},\left|GHZ\right\rangle }\left(\xi_{k}^{\prime\prime\prime}\right)=\frac18\;.
\end{equation}

The quantum correlation function in the considered case is also nontrivial:
\begin{equation}
\begin{array}[c]{c}
\phi_{q}\left(\frac{1}{2},\frac{1}{2},\frac{1}{2}\right) =2,\\
\phi_{q}\left(\frac{1}{2},\frac{1}{2},-\frac{1}{2}\right) =0,\\
\phi_{q}\left(\frac{1}{2},-\frac{1}{2},\frac{1}{2}\right) =0,\\
\phi_{q}\left(\frac{1}{2},-\frac{1}{2},-\frac{1}{2}\right) =2,\\
\phi_{q}\left(-\frac{1}{2},\frac{1}{2},\frac{1}{2}\right) =0,\\
\phi_{q}\left(-\frac{1}{2},\frac{1}{2},-\frac{1}{2}\right) =2,\\
\phi_{q}\left(-\frac{1}{2},-\frac{1}{2},\frac{1}{2}\right) =2,\\
\phi_{q}\left(-\frac{1}{2},-\frac{1}{2},-\frac{1}{2}\right) =0.
\end{array}
\end{equation}
what indicates a nontrivial quantum correlations between the three observables $\widehat{S}_{x,1},\widehat{S}_{y,2},\widehat{S}_{y,3}$ at $\left|GHZ\right\rangle$. Nevertheless, the three observables are weakly correlated than those considered before.

\section{Quantum correlations at the W state}
The W state
\begin{equation}
\left|W\right\rangle =\frac1{\sqrt{3}}\left(  \left|011\right\rangle+\left|101\right\rangle +\left|110\right\rangle \right)
\end{equation}
introduced by D\"{u}rr, Vidal, and Cirac \cite{drull:three} (see also \cite{cabello:bell} and references quoted therein), is distinguished -- like the GHZ state -- by its specific properties which make it a useful object for both theoretical studies and practical applications. We will calculate quantum correlations at $\left|W\right\rangle $ of the same two families of one-qubit spin observables as discussed above.
\subsection{Case 1}

The outcome measure for the joint observable $E^{A}$ at the state $\left|W\right\rangle $ is determined by the probability function $P_{A,\left|W\right\rangle }$ which can be easily calculated. One gets:
\begin{equation}
\begin{array}[c]{c}
P_{A,\left| W\right\rangle }\left(\frac{1}{2},\frac{1}{2},\frac{1}{2}\right)=0,\\
P_{A,\left| W\right\rangle }\left(\frac{1}{2},\frac{1}{2},-\frac{1}{2}\right)=0,\\
P_{A,\left| W\right\rangle }\left(\frac{1}{2},-\frac{1}{2},\frac{1}{2}\right)=0,\\
P_{A,\left| W\right\rangle }\left(\frac{1}{2},-\frac{1}{2},-\frac{1}{2}\right)=\frac{1}{3},\\
P_{A,\left| W\right\rangle }\left(-\frac{1}{2},\frac{1}{2},\frac{1}{2}\right)=0,\\
P_{A,\left| W\right\rangle }\left(-\frac{1}{2},\frac{1}{2},-\frac{1}{2}\right)=\frac{1}{3},\\
P_{A,\left| W\right\rangle }\left(-\frac{1}{2},-\frac{1}{2},\frac{1}{2}\right)=\frac{1}{3},\\
P_{A,\left| W\right\rangle }\left(-\frac{1}{2},-\frac{1}{2},-\frac{1}{2}\right)=0\;.
\end{array}
\end{equation}

The outcome measures for those three observables $\widehat{S}_{z,1}$,$\widehat{S}_{z,2}$, $\widehat{S}_{z,3}$ at $\left|W\right\rangle $ are determined by the probability functions:
\begin{equation}
\begin{array}[c]{c}
P_{\widehat{S}_{z,i},\left|W\right\rangle }\left(\frac{1}{2}\right)=\frac{1}{3}\\
P_{\widehat{S}_{z,i},\left|W\right\rangle }\left(-\frac{1}{2}\right)=\frac23
\end{array}
\end{equation}
for $i=1,2,3$. Consequently, the product probability function
\begin{equation}
P_{\widehat{S}_{z,1},\left|W\right\rangle}\left(\xi_{i}^{\prime}\right)\cdot P_{\widehat{S}_{z,2},\left|W\right\rangle }\left(\xi_{j}^{\prime\prime}\right)\cdot P_{\widehat{S}_{z,3},\left|W\right\rangle}\left(\xi_{k}^{\prime\prime\prime}\right)
\end{equation}
is
\begin{equation}
\begin{array}[c]{c}
\left(\frac{1}{2},\frac{1}{2},\frac{1}{2}\right)  \rightarrow\frac{1}{27},\\
\left(\frac{1}{2},\frac{1}{2},-\frac{1}{2}\right)  \rightarrow\frac{2}{27},\\
\left(\frac{1}{2},-\frac{1}{2},\frac{1}{2}\right)  \rightarrow\frac{2}{27},\\
\left(\frac{1}{2},-\frac{1}{2},-\frac{1}{2}\right)  \rightarrow\frac{4}{27},\\
\left(-\frac{1}{2},\frac{1}{2},\frac{1}{2}\right)  \rightarrow\frac{2}{27},\\
\left(-\frac{1}{2},\frac{1}{2},-\frac{1}{2}\right)  \rightarrow\frac{4}{27},\\
\left(-\frac{1}{2},-\frac{1}{2},\frac{1}{2}\right)  \rightarrow\frac{4}{27},\\
\left(-\frac{1}{2},-\frac{1}{2},-\frac{1}{2}\right)  \rightarrow\frac{8}{27}.
\end{array}
\end{equation}

The quantum correlation function is:
\begin{equation}
\begin{array}[c]{c}
\phi_{q}\left(\frac{1}{2},\frac{1}{2},\frac{1}{2}\right)=0,\\
\phi_{q}\left(\frac{1}{2},\frac{1}{2},-\frac{1}{2}\right)=0,\\
\phi_{q}\left(\frac{1}{2},-\frac{1}{2},\frac{1}{2}\right)=0,\\
\phi_{q}\left(\frac{1}{2},-\frac{1}{2},-\frac{1}{2}\right)=2\frac{1}{4},\\
\phi_{q}\left(-\frac{1}{2},\frac{1}{2},\frac{1}{2}\right)=0,\\
\phi_{q}\left(-\frac{1}{2},\frac{1}{2},-\frac{1}{2}\right)=2\frac{1}{4},\\
\phi_{q}\left(-\frac{1}{2},-\frac{1}{2},\frac{1}{2}\right)=2\frac{1}{4},\\
\phi_{q}\left(-\frac{1}{2},-\frac{1}{2},-\frac{1}{2}\right)=0\;.
\end{array}
\end{equation}
We see that the observables $\widehat{S}_{z,1},$ $\widehat{S}_{z,2}$,$\widehat{S}_{z,3}$ are strongly quantum correlated at $\left|W\right\rangle$.
\subsection{Case 2}

The joint observable $E^{B}$ generates at $\left|W\right\rangle $ the following outcome probability function%
\begin{equation}
\begin{array}[c]{c}
P_{B,\left|W\right\rangle }\left(\frac{1}{2},\frac{1}{2},\frac{1}{2}\right)=\frac{5}{24}\;,\\
P_{B,\left|W\right\rangle }\left(\frac{1}{2},\frac{1}{2},-\frac{1}{2}\right)=\frac{1}{24}\;,\\
P_{B,\left|W\right\rangle }\left(\frac{1}{2},-\frac{1}{2},\frac{1}{2}\right)=\frac{1}{24}\;,\\
P_{B,\left|W\right\rangle }\left(\frac{1}{2},-\frac{1}{2},-\frac{1}{2}\right)=\frac5{24}\;,\\
P_{B,\left|W\right\rangle }\left(-\frac{1}{2},\frac{1}{2},\frac{1}{2}\right)=\frac5{24}\;,\\
P_{B,\left|W\right\rangle }\left(-\frac{1}{2},\frac{1}{2},-\frac{1}{2}\right)=\frac{1}{24}\;,\\
P_{B,\left|W\right\rangle }\left(-\frac{1}{2},-\frac{1}{2},\frac{1}{2}\right)=\frac{1}{24}\;,\\
P_{B,\left|W\right\rangle }\left(-\frac{1}{2},-\frac{1}{2},-\frac{1}{2}\right)=\frac5{24}\;.
\end{array}
\end{equation}
The outcome measures for those three observables $\widehat{S}_{x,1},\widehat {S}_{y,2},\widehat{S}_{y,3}$ at $\left|W\right\rangle $ are uniformly distributed, like:
\begin{equation}
\begin{array}[c]{c}
P_{\widehat{S}_{x,1},\left|W\right\rangle }\left(\frac{1}{2}\right)=\frac{1}{2},\\
P_{\widehat{S}_{x,1},\left|W\right\rangle }\left(-\frac{1}{2}\right)=\frac{1}{2}.
\end{array}
\end{equation}
Hence, the product probability function is constant:
\begin{equation}
P_{\widehat{S}_{z,1},\left|W\right\rangle}\left(\xi_{i}^{\prime}\right)
\cdot P_{\widehat{S}_{z,2},\left|W\right\rangle}\left(\xi_{j}^{\prime\prime}\right)\cdot P_{\widehat{S}_{z,3},\left|W\right\rangle}\left(\xi_{k}^{\prime\prime\prime}\right)=\frac{1}8.
\end{equation}
The quantum correlation function in the considered case is also nontrivial:
\begin{equation}
\begin{array}[c]{c}
\phi_{q}\left(\frac{1}{2},\frac{1}{2},\frac{1}{2}\right)=\frac53\;,\\
\phi_{q}\left(\frac{1}{2},\frac{1}{2},-\frac{1}{2}\right)=\frac{1}{3}\;,\\
\phi_{q}\left(\frac{1}{2},-\frac{1}{2},\frac{1}{2}\right)=\frac{1}{3}\;,\\
\phi_{q}\left(\frac{1}{2},-\frac{1}{2},-\frac{1}{2}\right)=\frac53\;,\\
\phi_{q}\left(-\frac{1}{2},\frac{1}{2},\frac{1}{2}\right)=\frac53\;,\\
\phi_{q}\left(-\frac{1}{2},\frac{1}{2},-\frac{1}{2}\right)=\frac{1}{3}\;,\\
\phi_{q}\left(-\frac{1}{2},-\frac{1}{2},\frac{1}{2}\right)=\frac{1}{3}\;,\\
\phi_{q}\left(-\frac{1}{2},-\frac{1}{2},-\frac{1}{2}\right)=\frac53\;.
\end{array}
\end{equation}
what indicates a nontrivial quantum correlation between the three observables$\widehat{S}_{x,1},\widehat{S}_{y,2},\widehat{S}_{y,3}$ at $\left|W\right\rangle $. Nevertheless, the three observables are weaker correlated than these considered as Case 1.

\section{Bipartite entanglement at the tripartite states}
\subsection{GHZ state}

The three spin observables $\widehat{S}_{z,1},$ $\widehat{S}_{z,2}$, $\widehat{S}_{z,3}$ admit, besides the ''tripartite'' joint observable $E^{A},$ Eq.~(\ref{A}), also ''bipartite'' joint observables. Thus, for instance, the quantum joint observable $E^{D}$ for $\widehat{S}_{z,1}$ and $\widehat{S}_{z,2}$ is represented by the following PV function:
\begin{equation}
\begin{array}[c]{c}
E^{D}\left(\frac{1}{2},\frac{1}{2}\right)=\left|0\right\rangle \left\langle
0\right|  \otimes\left|0\right\rangle \left\langle 0\right|  \otimes\widehat{1}\\
E^{D}\left(\frac{1}{2},-\frac{1}{2}\right)=\left|0\right\rangle \left\langle
0\right|  \otimes\left|1\right\rangle \left\langle 1\right|\otimes\widehat{1}\\
E^{D}\left(-\frac{1}{2},\frac{1}{2}\right)=\left|1\right\rangle \left\langle
1\right|\otimes\left|0\right\rangle \left\langle 0\right|\otimes\widehat{1}\\
E^{D}\left(-\frac{1}{2},-\frac{1}{2}\right)=\left|1\right\rangle \left\langle
1\right|\otimes\left|1\right\rangle \left\langle 1\right|\otimes\widehat{1}
\end{array}
\label{D}
\end{equation}
The outcome measure for those joint observable $E^{D}$ at the state $\left|GHZ\right\rangle $ is now defined by the probability function:
\begin{equation}
\begin{array}[c]{c}
P_{D,\left|GHZ\right\rangle }\left(\frac{1}{2},\frac{1}{2}\right)=\frac{1}{2}\\
P_{D,\left|GHZ\right\rangle }\left(\frac{1}{2},-\frac{1}{2}\right)=0\\
P_{D,\left|GHZ\right\rangle }\left(-\frac{1}{2},\frac{1}{2}\right)=0\\
P_{D,\left|GHZ\right\rangle }\left(-\frac{1}{2},-\frac{1}{2}\right)=\frac{1}{2}
\end{array}
\end{equation}
The outcome measures for observables $\widehat{S}_{z,1}$and $\widehat{S}_{z,2}$ at $\left|GHZ\right\rangle $ have been calculated above; the product probability function is constant:
\begin{equation}
P_{\widehat{S}_{z,1},\left|GHZ\right\rangle }\left(\xi_{i}^{\prime}\right)\cdot P_{\widehat{S}_{z,2},\left|GHZ\right\rangle }\left(\xi_{j}^{\prime\prime}\right)=\frac{1}{4},
\end{equation}
so generates a uniformly distributed probability measure. We obtain the following quantum correlation function:
\begin{equation}
\begin{array}[c]{c}
\phi_{q}\left(\frac{1}{2},\frac{1}{2}\right)=2,\\
\phi_{q}\left(\frac{1}{2},-\frac{1}{2}\right)=0,\\
\phi_{q}\left(-\frac{1}{2},\frac{1}{2}\right)=0,\\
\phi_{q}\left(-\frac{1}{2},-\frac{1}{2}\right)=2,
\end{array}
\end{equation}
which shows that the $z$-th coordinates of spins of the 1sth and the 2nd qubit, $\widehat{S}_{z,1}$ and $\widehat{S}_{z,2},$ are strongly correlated at $\left|GHZ\right\rangle$. The strong quantum correlation between $\widehat{S}_{z,1}$ and $\widehat{S}_{z,2}$ apparently contradicts the known fact that reduced states of $\left|GHZ\right\rangle$ are separable (see \cite{drull:three}).

Consider now the same at the level of bipartite system of the 1sth and the 2nd qubits. The two comeasurable observables
\begin{equation}
\widehat{S}_{z,1}^{\prime}:=\widehat{s}_{z}\otimes\widehat{1}\ ,\ \widehat{S}_{z,2}^{\prime}:=\widehat{1}\otimes\widehat{s}_{z}
\end{equation}
have the joint observable
\begin{equation}
\begin{array}[c]{c}
E^{D^{\prime}}\left(\frac{1}{2},\frac{1}{2}\right)=\left|0\right\rangle
\left\langle 0\right|\otimes\left|0\right\rangle \left\langle 0\right|,\\
E^{D^{\prime}}\left(\frac{1}{2},-\frac{1}{2}\right)=\left|0\right\rangle
\left\langle 0\right|\otimes\left|1\right\rangle \left\langle 1\right|,\\
E^{D^{\prime}}\left(-\frac{1}{2},\frac{1}{2}\right)=\left|1\right\rangle
\left\langle 1\right|\otimes\left|0\right\rangle \left\langle 0\right|,\\
E^{D^{\prime}}\left(-\frac{1}{2},-\frac{1}{2}\right)=\left|1\right\rangle
\left\langle 1\right|\otimes\left|1\right\rangle \left\langle 1\right|
\end{array}
\label{D'}
\end{equation}

The reduced density operator
\begin{equation}
\widehat{\rho}^{\prime}:=\mathrm{Tr}_{3}\left|GHZ\right\rangle \left\langle GHZ\right|
\end{equation}
decomposes into
\begin{equation}
\widehat{\rho}^{\prime}=\frac{1}{2}\left(  \left|00\right\rangle \left\langle
00\right|  +\left|11\right\rangle \left\langle 11\right|\right)
\end{equation}
so it is a separable mixed state. The reduced state $\widehat{\rho}^{\prime}$ produces the same outcome at the two-qubit joint observable $E^{D^{\prime}}$ as the three-qubit state $\left|GHZ\right\rangle $ does at the joint observable $E^{D}$. Indeed, the probability function generating the outcome measure for $E^{D^{\prime}}$ at $\widehat{\rho}^{\prime}$ is:
\begin{equation}
\begin{array}[c]{c}
P_{D^{\prime},\widehat{\rho}^{\prime}}\left(\frac{1}{2},\frac{1}{2}\right)=\frac{1}{2}\\
P_{D^{\prime},\widehat{\rho}^{\prime}}\left(\frac{1}{2},-\frac{1}{2}\right)=0\\
P_{D^{\prime},\widehat{\rho}^{\prime}}\left(-\frac{1}{2},\frac{1}{2}\right)=0\\
P_{D^{\prime},\widehat{\rho}^{\prime}}\left(-\frac{1}{2},-\frac{1}{2}\right)=\frac{1}{2}
\end{array}
\end{equation}
This time, however, the probability function $P_{D^{\prime},\widehat{\rho}^{\prime}},$ although identical with $P_{D,\left|GHZ\right\rangle},$ is generated by the mixed separable state $\widehat{\rho}^{\prime},$ hence should
show no quantum correlations between $\widehat{S}_{z,1}^{\prime}$ and$\;\widehat{S}_{z,2}^{\prime}$ (compare \cite{beltrametti:entanglement}).

It is easy to find, however, that the reduced state $\widehat{\rho}^{\prime}$ could have a~different statistical content, as it admits the decomposition:
\begin{equation}
\widehat{\rho}^{\prime}=\frac{1}{2}\left(\left|B_{1}\right\rangle \left\langle B_{1}\right|+\left|B_{2}\right\rangle \left\langle B_{2}\right|\right),
\end{equation}
where
\begin{equation}
\left|B_{1}\right\rangle
=\frac{1}{\sqrt{2}}\left(\left|00\right\rangle
+\left|11\right\rangle \right)  \;,\;\left|  B_{2}\right\rangle
=\frac{1}{\sqrt{2}}\left(  \left|00\right\rangle -\left|11\right\rangle
\right)
\end{equation}
are two states of the Bell base. According to a general result of \cite{beltrametti:entanglement},
this statistical content of $\widehat{\rho}^{\prime}$ implies that the two observables $\widehat{S}_{z,1}^{\prime}$,$\;\widehat{S}_{z,2}^{\prime}$ are quantum correlated without a classical correlation. Thus, we see that the
popular claim that reduced states of $\left|GHZ\right\rangle $ show no entanglement is conditioned on the statistical content of the reduced state.
\subsection{W state.}
We will also check how the same problem will look at the W state. The outcome measure for the joint observable $E^{D}$ at the state $\left|W\right\rangle $ is now defined by the probability function:
\begin{equation}
\begin{array}[c]{c}
P_{D,\left|W\right\rangle }\left(\frac{1}{2},\frac{1}{2}\right)=0,\\
P_{D,\left|W\right\rangle }\left(\frac{1}{2},-\frac{1}{2}\right)=\frac{1}{3},\\
P_{D,\left|W\right\rangle }\left(-\frac{1}{2},\frac{1}{2}\right)=\frac{1}{3},\\
P_{D,\left|W\right\rangle }\left(-\frac{1}{2},-\frac{1}{2}\right)=\frac{1}{3}.
\end{array}
\end{equation}
The outcome measures for observables $\widehat{S}_{z,1}$and $\widehat{S}_{z,2} $ at $\left|W\right\rangle $ have been calculated above; the product probability function $P_{\widehat{S}_{z,1},\left|W\right\rangle }\left(\xi_{i}^{\prime}\right)  \cdot P_{\widehat{S}_{z,2},\left|W\right\rangle}\left(  \xi_{j}^{\prime\prime}\right)  $ is:
\begin{equation}
\begin{array}[c]{c}
\left(\frac{1}{2},\frac{1}{2}\right)  \rightarrow\frac{1}{9},\\
\left(\frac{1}{2},-\frac{1}{2}\right)  \rightarrow\frac{2}{9},\\
\left(-\frac{1}{2},\frac{1}{2}\right)  \rightarrow\frac{2}{9},\\
\left(-\frac{1}{2},-\frac{1}{2}\right)  \rightarrow\frac{4}{9}.
\end{array}
\end{equation}
We obtain the following quantum correlation function:
\begin{equation}
\begin{array}[c]{c}
\phi_{q}\left(\frac{1}{2},\frac{1}{2}\right)=0\\
\phi_{q}\left(\frac{1}{2},-\frac{1}{2}\right)=\frac32\\
\phi_{q}\left(-\frac{1}{2},\frac{1}{2}\right)=\frac32\\
\phi_{q}\left(-\frac{1}{2},-\frac{1}{2}\right)=\frac34
\end{array}
\end{equation}
which shows that the $z$-th coordinates of spins of the 1sth and the 2nd  qubit, $\widehat{S}_{z,1}$ and $\widehat{S}_{z,2},$ are quantum correlated at the state $\left|W\right\rangle$. The reduced density operator is
\begin{equation}
\widehat{\rho}^{\prime\prime}:=\mathrm{Tr}_{3}\left|W\right\rangle
\left\langle W\right|
\end{equation}
can be decomposed into:
\begin{equation}
\widehat{\rho}^{\prime\prime}=\frac23\left|  B_{3}\right\rangle \left\langle
B_{3}\right|  +\frac{1}{3}\left|11\right\rangle \left\langle 11\right|\;,
\end{equation}
where $\left|  B_{3}\right\rangle =\frac{1}{\sqrt{2}}\left(  \left|
01\right\rangle +\left|10\right\rangle \right)  $ is one of the Bell states.
The outcome probability measure for the joint observable $E^{D^{\prime}}$ at $\widehat{\rho}^{\prime\prime}$ is then provided by the probability function
\begin{equation}
\begin{array}[c]{c}
P_{D^{\prime},\widehat{\rho}^{\prime\prime}}\left(\frac{1}{2},\frac{1}{2}\right)=0,\\
P_{D^{\prime},\widehat{\rho}^{\prime\prime}}\left(\frac{1}{2},-\frac{1}{2}\right)=\frac{1}{3},\\
P_{D^{\prime},\widehat{\rho}^{\prime\prime}}\left(-\frac{1}{2},\frac{1}{2}\right)=\frac{1}{3},\\
P_{D^{\prime},\widehat{\rho}^{\prime\prime}}\left(-\frac{1}{2},-\frac{1}{2}\right)=\frac{1}{3}
\end{array}
\end{equation}
which, as it should be, equals $P_{D,\left|W\right\rangle}$. The probability functions generated by the two spin observables $\widehat{S}_{z,1}^{\prime}:=\widehat{s}_{z}\otimes\widehat{1}\;,\;\widehat{S}_{z,2}^{\prime}:=\widehat{1}\otimes\widehat{s}_{z}$ at the state $\widehat{\rho}^{\prime\prime}$ are:
\begin{equation}
\begin{array}[c]{c}
P_{\widehat{S}_{z,1}^{\prime},\widehat{\rho}^{\prime\prime}}\left(\frac{1}{2}\right)=\frac{1}{3},\\
P_{\widehat{S}_{z,1}^{\prime},\widehat{\rho}^{\prime\prime}}\left(-\frac{1}{2}\right)=\frac{2}{3},
\end{array}
\end{equation}
and
\begin{equation}
\begin{array}[c]{c}
P_{\widehat{S}_{z,2}^{\prime},\widehat{\rho}^{\prime\prime}}\left(\frac{1}{2}\right)=\frac{1}{3},\\
P_{\widehat{S}_{z,2}^{\prime},\widehat{\rho}^{\prime\prime}}\left(-\frac{1}{2}\right)=\frac{2}{3}
\end{array}
\end{equation}
so we get the same product probability function
\begin{equation}
\begin{array}[c]{c}
\left(\frac{1}{2},\frac{1}{2}\right)\rightarrow\frac{1}{9},\\
\left(\frac{1}{2},-\frac{1}{2}\right)\rightarrow\frac{2}{9},\\
\left(-\frac{1}{2},\frac{1}{2}\right)\rightarrow\frac{2}{9},\\
\left(-\frac{1}{2},-\frac{1}{2}\right)\rightarrow\frac{4}{9},
\end{array}
\end{equation}
and the same correlation function as for the state $\left|W\right\rangle$ considered above. We get
\begin{equation}
\begin{array}[c]{c}
\phi_{t}\left(\frac{1}{2},\frac{1}{2}\right)=0\\
\phi_{t}\left(\frac{1}{2},-\frac{1}{2}\right)=\frac{3}{2}\\
\phi_{t}\left(-\frac{1}{2},\frac{1}{2}\right)=\frac{3}{2}\\
\phi_{t}\left(-\frac{1}{2},-\frac{1}{2}\right)=\frac{3}{4}
\end{array}
\end{equation}
We should notice, however, that now the state in question is not pure, so we cannot \textit{a priori }exclude a classical correlation. The obtained correlation function describes total correlation, which includes both classical and quantum correlations; now we have to separate them.

According to considerations of Subsection 3.2, we have to calculate the sum probability function for $\widehat{S}_{z,1}^{\prime},\;\widehat{S}_{z,2}^{\prime}$ at the mixed state $\widehat{\rho}^{\prime\prime}=\frac23\left|B_{3}\right\rangle \left\langle B_{3}\right|  +\frac{1}{3}\left|11\right\rangle \left\langle 11\right|$. We get:
\begin{equation}
\begin{array}[c]{c}
\left(\frac{1}{2},\frac{1}{2}\right)\rightarrow\frac{1}{6},\\
\left(\frac{1}{2},-\frac{1}{2}\right)\rightarrow\frac{1}{6},\\
\left(-\frac{1}{2},\frac{1}{2}\right)\rightarrow\frac{1}{6},\\
\left(-\frac{1}{2},-\frac{1}{2}\right)\rightarrow\frac{1}{2},
\end{array}
\end{equation}
hence the classical correlation function is:
\begin{equation}
\begin{array}[c]{c}
\phi_{c}\left(\frac{1}{2},\frac{1}{2}\right)=\frac{3}{2},\\
\phi_{c}\left(\frac{1}{2},-\frac{1}{2}\right)=\frac34,\\
\phi_{c}\left(-\frac{1}{2},\frac{1}{2}\right)=\frac34,\\
\phi_{c}\left(-\frac{1}{2},-\frac{1}{2}\right)=\frac98.
\end{array}
\end{equation}
The quantum correlation function can be calculated directly, Eq.~(\ref{QCF}),or from the product rule, Eq.~(\ref{PR}). We get:
\begin{equation}
\begin{array}[c]{c}
\phi_{q}\left(\frac{1}{2},\frac{1}{2}\right)=0,\\
\phi_{q}\left(\frac{1}{2},-\frac{1}{2}\right)=2,\\
\phi_{q}\left(-\frac{1}{2},\frac{1}{2}\right)=2,\\
\phi_{q}\left(-\frac{1}{2},-\frac{1}{2}\right)=\frac{2}{3}\;.
\end{array}
\end{equation}
We see that classically the parallel directions of the two spins are correlated, whereas quantum correlations connect opposite directions of spins. The appearance of quantum correlation is in a qualitative agreement with the result of \cite{drull:three} stating that the reduced state of $\left| W\right\rangle $ shows entanglement.
\subsection{Comments}
Both examples show that the measurement of the joint observable $E^{D}$, Eq.~(\ref{D}), at a tripartite state is not equivalent to a measurement of the joint observable $E^{D^{\prime}}$, Eq.~(\ref{D'}) at the corresponding reduced state, in spite of the known fact that both measurements produce the same outcomes. The deep difference lies in correlations: the total correlation does not change, but its separation on the classical and the quantum one is different for both measurements.

\section{Conclusions}
Probabilistic entanglement is based on simple assumptions and
describes correlations in general form. This approach allow us to
separate classical and quantum correlations and quantify amount of
those  in general quantum states. Thus probabilistic entanglement
can be used an entanglement (quantum correlation) measure.

\section{Acknowledgment}
This Supported by the Polish State Committee for Scientific
Research (KBN) under the Project Nr 7 T11C 017 21. J. A. M. would
like to thank J. S\l adkowski for help in preparation of this
paper.
\nocite{werner:open} \nocite{horodecki:uniq}
\nocite{zeilinger:quantcorrel} \nocite{greenberger:bell}
\nocite{mermin:mysteries} \nocite{mermin:correlations98}
\nocite{nelson:experimental} \nocite{parthasarathy:intoprob}
\nocite{plesch:entangled}

\end{document}